\documentclass[runningheads,a4paper]{llncs}

\title{FlexDM: Enabling robust and reliable parallel data mining using WEKA}
\author{Madison Flannery$^{1}$, David M Budden$^2$ and Alexandre Mendes$^3$}

\institute{$^1$ Melbourne Graduate School of Science, University of Melbourne, Parkville, VIC 3010, Australia \\
$^2$ Systems Biology Laboratory, University of Melbourne, Parkville, VIC 3010, Australia\\
$^3$ School of Electrical Engineering and Computer Science, The University of Newcastle, Callaghan, NSW 2308, Australia\\
        \email{Alexandre.Mendes@newcastle.edu.au}}

\date{}

\usepackage[table]{xcolor}
\usepackage{times}
\usepackage{url}

\usepackage{graphicx}
\usepackage{amsmath}

\begin{document}

\maketitle

\begin{abstract}
Performing massive data mining experiments with multiple datasets and methods is a common task faced by most bioinformatics and computational biology laboratories. WEKA is a machine learning package designed to facilitate this task by providing tools that allow researchers to select from several classification methods and specific test strategies. Despite its popularity, the current WEKA environment for batch experiments, namely \textsl{Experimenter}, has four limitations that impact its usability: the selection of value ranges for methods options lacks flexibility and is not intuitive; there is no support for parallelisation when running large-scale data mining tasks; the XML schema is difficult to read, necessitating the use of the Experimenter's graphical user interface for generation and modification; and robustness is limited by the fact that results are not saved until the last test has concluded.

FlexDM implements an interface to WEKA to run batch processing tasks in a simple and intuitive way. In a short and easy-to-understand XML file, one can define hundreds of tests to be performed on several datasets. FlexDM also allows those tests to be executed asynchronously in parallel to take advantage of multi-core processors, significantly increasing usability and productivity. Results are saved incrementally for better robustness and reliability.

FlexDM is implemented in Java and runs on Windows, Linux and OSX. As we encourage other researchers to explore and adopt our software, FlexDM is made available as a pre-configured bootable reference environment. All code, supporting documentation and usage examples are also available for download at \texttt{http://sourceforge.net/projects/flexdm}.
\end{abstract}

\section{Introduction}

Large-scale data analysis is an integral component of bioinformatics and computational biology research. Genome-based studies (\emph{e.g.} microarray and next generation sequencing) allow tens of thousands of gene expression levels to be quantified and analysed simultaneously (\cite{budden2014predictive,hurley2014nail}). In such situations, sophisticated statistical and data mining tools have to be used to fully understand the massive amounts of data generated. WEKA (\textsl{Waikato Environment for Knowledge Analysis}) is one of those tools. Designed for machine learning and data mining, WEKA is a widely-used gold standard framework that has been cited more than 7,000 times in scientific literature (\cite{hall2009weka}).

When a specific dataset is to be analysed, it is difficult to know \emph{a priori} which combination of classifier and hyperparameters will provide a better understanding of the data. Consequently, WEKA data mining experiments typically involve the comparison of multiple classifiers to identify which best approximates the unknown function relating input features (epitomised by the recent work of~\cite{marsden2013language}). To facilitate this process, WEKA provides a module called \textsl{Experimenter}. Even though the Experimenter partially automates some data mining tasks, we have identified four major limitations with its functionality, as described in the following section.
\section{Approach}
FlexDM addresses the following major limitations in the default WEKA Experimenter:

\begin{itemize}
\item \textbf{Limitations to value ranges for methods options}: WEKA allows the setting of value ranges for hyperparameters by using the meta-classifiers \texttt{CVParameterSelection} and \texttt{GridSearch}. However, they allow only one or two method options to be tested at a time, respectively. In addition, the setup of those meta-classifiers requires several nonintuitive steps. FlexDM imposes no such limitations; it allows any number of value ranges, and all combinations are tested in parallel.

\item \textbf{Lack of parallel processing}: WEKA does not inherently utilise parallel processing when executing data mining tasks. Given that all mainstream CPUs are multicore with hyper-threading, this represents a missed opportunity for increased performance and productivity. Parallel processing for WEKA has previously been achieved with other packages(\emph{e.g.} Weka-Parallel~\cite{celis2002weka} and Grid-enabled Weka~\cite{khoussainov2004grid}) and has been incorporated as part of the FlexDM functionality accordingly, using asynchronous parallel processing to achieve maximum performance.

\item \textbf{Reliance on the Experimenter GUI}: WEKA uses an XML schema to represent the set of experiments to be executed, which gives a high degree of flexibility and extensibility to the tool (\cite{achard2001xml,varde2010xml}). However, the schema used by WEKA is very difficult to understand and modify without using the software's GUI. FlexDM uses a simplified XML schema with improved readability, allowing users to rapidly change test specifications without the need for a GUI.

\item \textbf{Lack of robustness}: The Experimenter runs all tests in sequence, but results are saved only after the last one has concluded. In real world applications, such tests often take days to weeks of processing, which means that a hardware failure will cause the loss of all intermediate results. FlexDM saves results incrementally as each job concludes, allowing the user to restart from the last task successfully completed.
\end{itemize}


\section{Example Application}

The FlexDM XML schema is intuitive and easily modified. Figure~\ref{fig:XMLfile} shows an example input file for the execution of two classifiers (\textsl{J48} and \textsl{PART}), with one hyperparameter (\textsl{confidence factor} - C) varying from 0.1 to 1.0 in increments of 0.1. This input file specifies leave-one-out cross-validation on the dataset ``\textsl{health.arff}". The confusion matrix will be saved in each results file. FlexDM also creates a \textit{summary file} containing the classification accuracy for each test, allowing easy identification of the best performing classifier.

The compact representation of FlexDM is enabled by the assumption that any information not specified (\emph{e.g.} hyperparameters) defaults to WEKA's default values. The file in Figure~\ref{fig:XMLfile} is 11 lines long and has 321 characters (excluding spaces), 10x more compact than the WEKA equivalent (available on the supplementary material webpage).

\begin{figure}[h]
\texttt{{\textless !DOCTYPE flexdm SYSTEM "flexdm.dtd"\textgreater
\\[-0.14cm]
    \textless flexdm\textgreater \\[-0.14cm]
        \hspace*{0.5cm}\textless dataset name="health.arff" test="leavexval" results = "matrix"\textgreater \\[-0.14cm]
		\hspace*{1.0cm}\textless classifier name="weka.classifiers.trees.J48"\textgreater \\[-0.14cm]
		\hspace*{1.5cm}\textless parameter name="-C" value="[0.1:0.1:1.0]" /\textgreater \\[-0.14cm]
		\hspace*{1.0cm}\textless /classifier\textgreater \\[-0.14cm]
		\hspace*{1.0cm}\textless classifier name="weka.classifiers.rules.PART"\textgreater \\[-0.14cm]
		\hspace*{1.5cm}\textless parameter name="-C" value="[0.1:0.1:1.0]" /\textgreater \\[-0.14cm]
		\hspace*{1.0cm}\textless /classifier\textgreater \\[-0.14cm]
		\hspace*{0.5cm}\textless /dataset\textgreater \\[-0.14cm]
		\textless /flexdm\textgreater \\[-0.5cm]}}
		
\caption{Example of XML input file for FlexDM. The schema used is intuitive and easily modified. The option ``C" is given a range [0.1:0.1:1.0], resulting in 10 independent tests for each classifier with C = 0.1, 0.2, ..., 1.0. Comparatively, the equivalent WEKA XML file has 131 lines and 6,367 characters.}
\label{fig:XMLfile}
\end{figure}

\begin{figure}[!t]
\begin{center}
\includegraphics[width=10cm]{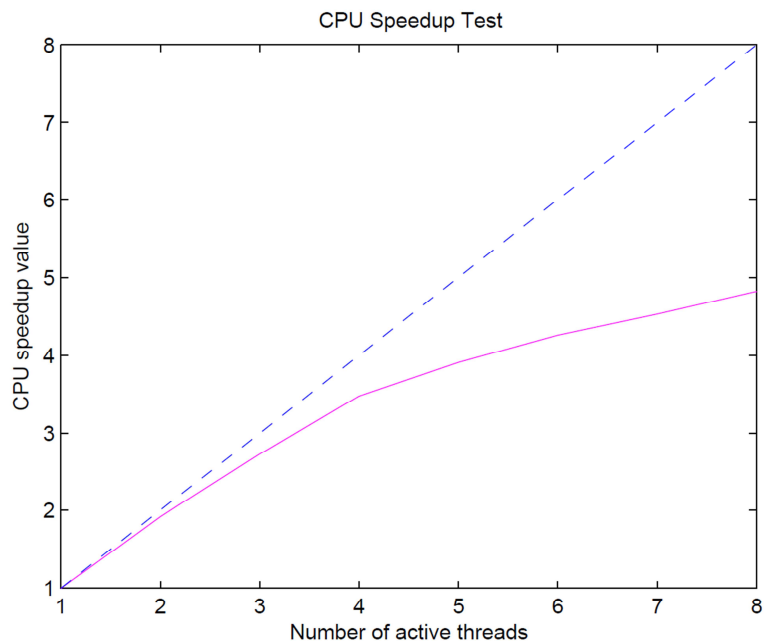}
\end{center}
\caption{Speedup generated by FlexDM's asynchronous parallel processing (pink) compared to theoretical maximum (blue), using a quad-core CPU.}
\vspace{-0.3cm}
\label{fig:Speedup}
\end{figure}

Another feature of FlexDM is its asynchronous parallel processing. If an input XML file specifies more than one classifier, the software runs them in parallel using exactly (n-1) logical core units present in the host computer (one core is left available for other activities). In Figure~\ref{fig:Speedup}, we show the speedup obtained with the XML test file from Figure~\ref{fig:XMLfile}. The test was carried out on an Intel i7 processor (quad-core CPU with hyper-threading). Note the quasi-linear speedup up until four threads, followed by a less steep speedup curve once hyper-threading is used.


\section{Implementation}
As we encourage other researchers to explore and adopt our software, FlexDM is implemented using open-source software and made available as a pre-configured bootable virtual environment using the approach described by~\cite{hurley2014virtual}. This environment was created using a minimal installation of Lubuntu 13.10; a lightweight Linux distribution which supports all the tools required. OpenJDK version 2.4.7 and WEKA version 3.6.10 were installed, along with the core set of packages and utilities required to reproduce our usage examples. Full configuration details for the reference environment are available in the Supplementary Materials. Alternatively, all data and scripts are available online at \texttt{http://sourceforge.net/projects/flexdm}.

\section{Conclusion}
This paper presents a software tool for robust and reliable parallel data mining using WEKA, FlexDM, which addresses four major limitations of the default WEKA Experimenter. The use of a simplified XML schema also makes FlexDM far easier to learn than the Experimenter. This tool can significantly improve productivity for any research group that frequently performs large-scale data mining or machine learning tasks, such as those commonplace in the bioinformatics and computational biology domain.



\bibliographystyle{splncs}
\bibliography{flannery_2014_flexdm}






\end{document}